\documentclass[twocolumn,showpacs,prl,preprintnumbers,floatfix]{revtex4}

\usepackage{graphics}
\usepackage{mathrsfs}
\usepackage{amssymb}
\usepackage{amsmath}
\usepackage{graphicx,amssymb}
\usepackage{setspace}
\usepackage{dcolumn}
\usepackage{bm}
\usepackage{flafter}
\usepackage{subfigure}
\usepackage{times}
\usepackage{xspace}
\usepackage{wasysym}
\usepackage{ulem}
\usepackage{soul}
\usepackage{color}

\begin{document}

\preprint{APS/}
\title{\bf Record-breaking statistics for random walks in the presence of measurement error and noise}


\author{Yaniv Edery$^1$, Alexander B. Kostinski$^2$, Satya N. Majumdar$^3$ and Brian Berkowitz$^1$}

\affiliation{1. Department of Environmental Sciences and Energy
Research, Weizmann Institute of Science, 76100 Rehovot, Israel \\
2. Department of Physics, Michigan Technological University,1400
Townsend Drive, Houghton, MI 49931 USA \\
3. Laboratoire de Physique Th{\'e}orique et Mod{\`e}les
Statistiques (UMR 8626 du CNRS), Universit{\'e} Paris-Sud,
B{\^a}timent 100, 91405 Orsay Cedex, France}

\begin{abstract}
We address the question of distance record-setting by a random
walker in the presence of measurement error, $\delta$, and
additive noise, $\gamma$ and show that the mean number of (upper)
records up to $n$ steps still grows universally as $\langle
R_n\rangle \sim n^{1/2}$ for large $n$ for all jump distributions,
including L\'evy flights, and for all $\delta$ and $\gamma$.  In
contrast to the universal growth exponent of $1/2$, the pace of
record setting, measured by the pre-factor of $n^{1/2}$, depends
on $\delta$ and $\gamma$. In the absence of noise ($\gamma=0$),
the pre-factor $S(\delta)$ is evaluated explicitly for arbitrary
jump distributions and it decreases monotonically with increasing
$\delta$ whereas, in case of perfect measurement $(\delta=0)$, the
corresponding pre-factor $T(\gamma)$ increases with $\gamma$. Our
analytical results are supported by extensive numerical
simulations and qualitatively similar results are found in two and
three dimensions.

\end{abstract}


\maketitle

\par
An upper record (record, for short) occurs at step $n$ in a time
series if the $n$-th entry exceeds all previous entries. The
statistics of record breaking events in a discrete-time series
with independent and identically distributed (i.i.d) entries have
been studied in statistics and mathematics literature for a long
time~\cite{FS54,Arnold,Nevzorov}. Recent years have seen a
resurgence of interest in record statistics, which play a major
role in the analysis of time series in a number of diverse
contexts, including
sports~\cite{Glick78,Gembris2002,Gembris2007,BRV07}, biological
evolution models~\cite{KJ05,Krug07}, theory of spin-glasses
\cite{Oliveira2005,Sibani2006}, models of growing
networks\cite{GL08}, analysis of climate
data~\cite{Matalas97,Redner2006,Meehl2009,WK10,AK11}, and quantum
chaos~\cite{SLJ13}. The quantity of central interest is the mean
number of records $\langle R_n\rangle$ up to step $n$. For a time
series with i.i.d entries, a striking universal result is that
$\langle R_n\rangle \sim \ln n$ for large $n$~\cite{FS54},
independent of the distribution of the individual entries.
However, this universal logarithmic growth breaks down when the
entries of the time series are {\it strongly correlated}, the
simplest example being the case of a random walk where the entries
of the time series represent the positions of the walker at
discrete time steps.

\par
While the subject of random walks has an enormous range of
applications well beyond the original context of diffusion and
Brownian motion, its exploration in terms of record setting is
relatively recent. The basic question is: how often does a random
walker, moving in continuous space by jumping a random distance at
each discrete time step, set a distance record, i.e., advance
farther from the origin than at all prior steps? In other words,
how does the mean number of such record-setting events grow with
the number of steps? This is a very natural question in many
different contexts, such as in the evolution of stock
prices~\cite{Williams06,Yor2000} or in queueing
theory~\cite{Asmussen}. In the one-dimensional case, with pure
diffusion but no overall drift, a universally valid result was
found in~\cite{MZ08} for the mean of the upper record-setting
events $\langle R_n\rangle$, namely, that it scales as
$(2/\sqrt{\pi})\, n^{1/2}$ for large $n$, where $n$ is the number
of steps, regardless of the length distribution of jumps (e.g.,
holds even for L\'evy flights). This square root growth of the
mean record number was also found numerically in two and three
dimensions, and considering a drift, an abrupt shift in the
scaling exponent from $1/2$ to $1$ was identified~\cite{EKB11}.
Exact analytical results were also found in one dimension for a
random walker with arbitrary drift~\cite{PLDW09,MSW12}, for a
continuous time random walker~\cite{SS} and for multiple random
walkers~\cite{WMS12}. In the latter case, the theoretical results
were in good agreement with an analysis of multiple stocks from
the Standard \& Poors 500 index~\cite{WMS12}.

\par
However, to apply the above results to interpretation of
\textit{real} experiments, one needs to re-examine the notion of a
record because the phrase ``advance farther from the origin than
at all prior time steps'' requires closer examination. Why?
Because all \textit{real} measurements involve instrument error,
$\delta$, and noise, $\gamma$, is unavoidable. For instance,
$\delta$ can be the assurance limit of the detector while $\gamma$
can describe white noise from an instrument reading. Ties become
possible because of the ``fuzziness'', as discussed, for example,
in~\cite{AK10,AK11,WVRK12}.  Hence, the question arises: how does
the presence of measurement error $\delta$ or noise $\gamma$
affect the growth of record number and the associated
record-setting pace? Related questions were raised in the
statistics literature, e.g., in terms of $\delta$-exceedance
records~\cite{BBP96,GJS12} and very recently in the physics
literature~\cite{WVRK12}, but asymptotic results are available
only for time series with i.i.d entries. To the best of our
knowledge, the question has never been raised in the context of
correlated entries such as random walks.  For example, does the
$\langle R_n\rangle \sim n^{1/2}$ scaling persist despite the
presence of $\delta$ or $\gamma$ and for various jump length
distributions? If so, how is the pre-factor affected? We address
these questions here using exact calculations and detailed Monte
Carlo simulations. As a preview, our major finding is the
decoupling between the growth exponent which remains universal and
the pre-factor which carries the ``burden'' of finite precision
and noise.

\par
Rather than working with absolute values, we define a
``one-sided'' record (only positive maxima are considered) so that
the $i$-th entry in a time series, $x_{i}$, is a record-breaking
event (record, for short) if it exceeds all previous values in the
sequence, i.e., if $x_{i} >$ max $(x_{1}, x_{2}, \ldots,
x_{i-1})$.  We shall henceforth interpret $x_{i}$ as the distance
of the random walker from the origin at the $i$-th time step.
However, because of the presence of (a fixed) measurement error,
$\delta$, we shall now define $x_{i}$ to be a record-breaking
event ($\delta$-record, for short) only if it exceeds all previous
values in the sequence by, \textit{at least}, $\delta$. Similarly,
accounting for measurement noise, $x_{i}$ is a record-breaking
event if, with the addition of $\gamma$ (white noise), it exceeds
all previous values in the sequence.  A subtlety is that in the
presence of error, a record can be defined as being larger -- by
the amount of the error -- than the last record, or than the last
maximum, the two being identical in the absence of error. In the
analysis below, we enumerate records larger than the previous
maximum, as it is more amenable to theoretical development as we
show below.

\par
We focus first on the influence of measurement error $\delta$.
Consider a discrete-time sequence $\{x_0=0,x_1,x_2,\ldots,\}$,
representing the position of a one-dimensional random walker
starting at the origin $x_0=0$. The position $x_m$ at step $m$ is
a continuous stochastic variable that evolves via the Markov rule,
$x_m= x_{m-1}+\eta_m$ where $\eta_m$ represents the jump at step
$m$. The noise variables $\eta_m$'s are independent and
identically distributed random variables, each drawn from a
symmetric and continuous jump density $f(\eta)$. Note that
although $\eta_m$'s are uncorrelated, $x_m$'s are correlated
random variables. We are interested in the statistics of the
number of records $R_n$ up to step $n$. A record occurs at step
$m$ if $x_m-\delta \ge x_k$ for all $k=0,1,2,\ldots, (m-1)$ where
$\delta\ge 0$ represents the measurement error. For $\delta=0$,
the statistics of $R_n$ are known to be universal, i.e.,
independent of the jump density $f(\eta)$ \cite{MZ08}. For
instance, the mean record number $\langle R_n\rangle$ up to step
$n$ is given by the expression \cite{MZ08}
\begin{equation}
\langle R_n\rangle= (2n+1)\,{2n\choose n}\,2^{-2n}
\xrightarrow[n\to \infty]{} \frac{2}{{\pi}^{1/2}}\, n^{1/2} \, .
\label{mean_record.1}
\end{equation}

\par
We now proceed to examine how $\langle R_n\rangle$ is affected by
the measurement error $\delta$. Define an indicator
$\sigma_m=\{1,0\}$ at each step $m$ so that $\sigma_m=1$ if a
record occurs at step $m$ and is $0$ otherwise. We call $x_0=0$ a
record, i.e., $\sigma_0=1$. Then evidently the number of records
$R_n$ up to step $n$ in a given sequence is $R_n= \sum_{m=0}^n
\sigma_m$. Next, we average this expression over different
histories. Because $\sigma_m$ is a binary $\{1,0\}$ variable, its
average $\langle \sigma_m\rangle$ is just the probability that a
record occurs at step $m$. Hence,
\begin{equation}
\langle R_n \rangle = \sum_{m=0}^n \langle \sigma_m\rangle=
\sum_{m=0}^n r_m(\delta), \label{rnavg.1}
\end{equation}
where $r_m(\delta)$ denotes the record rate, i.e., the probability
that a record occurs at step $m$. By definition, $r_0=1$. Hence,
\begin{equation}
r_m(\delta)= {\rm Prob}\left[x_m-\delta\ge {\rm
max}\left[0,x_1,x_2,\ldots, x_{m-1}\right]\right]\, . \label{rm.1}
\end{equation}
Thus, $r_m(\delta)$ is the probability of the event that the
random walker, starting at the origin, reaches $x_m$ at step $m$,
while staying below $x_m-\delta$ at all intermediate steps between
$0$ and $m$, where one needs to finally integrate over all $x_m\ge
\delta$. To compute this probability, it is convenient to change
variables $y_{k}= x_m-x_{m-k}$, i.e., observe the sequence
$\{y_k\}$ with respect to the last position and measure time
backwards. Then, $r_m(\delta)$ is the probability that the new
walker $y_k$, starting at the new origin at $k=0$, makes a jump
$\ge \delta$ at the first step and then subsequently up to $m$
steps stays above $\delta$, i.e.,
\begin{equation}
r_m(\delta)= {\rm Prob}\left[y_1\ge \delta, y_2\ge \delta,\ldots,
y_m\ge \delta|y_0=0\right]  . \label{yk.1}
\end{equation}

\par
To compute the probability $r_m(\delta)$ in (\ref{yk.1}), we note
that in the first step, the walker jumps to $y_1= z+\delta$ from
$y_0=0$ where $z\ge 0$ and subsequently up to $(m-1)$ steps it
stays above the level $\delta$. Writing $y_k=z_k+\delta$, we can
re-express $r_m(\delta)$ as
\begin{equation}
r_m(\delta)= \int_{0}^{\infty} f(z+\delta)\, q_{m-1}(z) dz
\label{yk.2}
\end{equation}
where $q_n(z)$ is the probability that a random walker, starting
initially at $z$, stays positive up to $n$ steps. This persistence
probability $q_n(z)$ has been thoroughly studied in the literature
for random walks (for a review, see~\cite{SMreview}) with
arbitrary jump density $f(\eta)$ and a general expression for its
Laplace transform is known as the Pollaczek-Spitzer formula
\cite{Pollaczek,Spitzer}. It states that
\begin{equation}
\int_0^{\infty} dz\, e^{-\lambda z}\, \sum_{n=0}^{\infty} s^n
q_n(z) = \frac{1}{\lambda \sqrt{1-s}} \phi(s,\lambda) \label{ps1}
\end{equation}
where
\begin{equation}
\phi(s,\lambda)= \exp\left[-\frac{\lambda}{\pi} \int_0^{\infty}
\frac{\ln \left(1-s {\hat f}(k)\right)}{\lambda^2+k^2}\, dk
\right]
\end{equation}
and ${\hat f}(k)=\int_{\infty}^{\infty} f(\eta)\,
e^{i\,k\eta}\,d\eta$ is the Fourier transform of the jump density
$f(\eta)$. Note that when $\delta\to 0$, the integral in
(\ref{yk.2}) is just $q_m(0)$. Thus $r_m(0)=q_m(0)$. From the
Pollaczek-Spitzer formula in (\ref{ps1}), one can show
\cite{SMreview} that $\sum_{m=0}^{\infty} q_m(0) s^m=
1/\sqrt{1-s}$, independent of the jump density. This is the
celebrated Sparre Andersen theorem \cite{SA} and when inverted it
simply gives $q_m(0)= {{2m}\choose m}\,{2^{-2m}}$. When
substituted back in (\ref{rnavg.1}), it then provides the
universal result~\cite{MZ08} in (\ref{mean_record.1}).

\par
However, we are interested in the case of $\delta>0$. To compute
$r_m(\delta)$ for large $m$ in (\ref{yk.2}), we need to know the
large $m$ behavior of $q_m(z)$ for a fixed $z>0$. This can be
extracted by analyzing (\ref{ps1}) near $s=1$. One finds that the
leading order behavior of the right hand side of (\ref{ps1}) near
$s=1$ is simply $[\phi(1,\lambda)/\lambda] (1-s)^{-1/2}$. This
means that $q_n(z)$ for large $n$, with fixed $z$, must behave
like $q_n(z)\approx h(z)/\sqrt{\pi\, n}$. Substituting this on the
left side of (\ref{ps1}) and analyzing  the leading behavior near
$s=1$ shows that the left hand side of (\ref{ps1}), near $s=1$,
behaves as ${\tilde h}(\lambda) (1-s)^{-1/2}$, where ${\tilde
h}(\lambda)= \int_0^{\infty} h(z) e^{-\lambda z} dz$ is the
Laplace transform of $h(z)$. Comparing the left and right sides of
(\ref{ps1}), we obtain, for large $n$

\begin{equation}
q_n(z) \approx \frac{h(z)}{\sqrt{\pi n} }\quad {\rm with}\quad
{\tilde h}(\lambda)=\int_0^{\infty} h(z)\, e^{-\lambda z}\, dz=
\frac{1}{\lambda}\phi(1,\lambda) \label{qnz.1}
\end{equation}
where $\phi(1,\lambda)$ can be read off (\ref{ps1}) as
\begin{equation}
\phi(1,\lambda)= \exp\left[-\frac{\lambda}{\pi} \int_0^{\infty}
\frac{\ln \left(1- {\hat f}(k)\right)}{\lambda^2+k^2}\, dk
\right]\, . \label{phi1}
\end{equation}
Substituting the asymptotic behavior of $q_n(z)$ from
(\ref{qnz.1}) in (\ref{yk.2}), we obtain, for large $m$,
%
\begin{equation}
r_m(\delta)\approx \frac{U(\delta)}{\sqrt{\pi m}}, \quad
U(\delta)= \int_0^{\infty} dz f(z+\delta) h(z)\, . \label{rm.3}
\end{equation}

\par
Finally, substituting this asymptotic behavior of the record rate
$r_m(\delta)$ in Eq. (\ref{rnavg.1}) and performing the sum for
large $n$, we find that the mean number of records for large $n$
is given by
%
\begin{equation}
\langle R_n\rangle \xrightarrow[n\to \infty]{} S(\delta)\,
{n}^{1/2}, \quad S(\delta)= \frac{2}{\sqrt{\pi}} \int_0^{\infty}
f(z+\delta) h(z) dz\, . \label{rnavg.2}
\end{equation}
This is the main exact result: for an arbitrary jump density
$f(\eta)$, the mean record number grows universally as $n^{1/2}$
for large $n$ (as in the $\delta=0$ case), while the pre-factor
$S(\delta)$ depends on $\delta$ and does so non-universally
insofar as its expression depends explicitly on the jump density
$f(\eta)$.


Although we have an exact expression for $S(\delta)$ for arbitrary
$f(\eta)$, its explicit evaluation for all $\delta$ is difficult.
For instance, to compute it explicitly for arbitrary jump density
$f(\eta)$, we need to first compute its Fourier transform ${\hat
f}(k)$, evaluate $\phi(1,\lambda)/\lambda$ from (\ref{phi1}), then
invert the Laplace transform (\ref{qnz.1}) to obtain $h(z)$ and
finally perform the integral in (\ref{rnavg.2}) to determine the
amplitude $S(\delta)$.

\par
For the special case of an exponential jump density,
$f(\eta)=(b/2) \exp[-b\,|\eta|]$, it is possible to evaluate the
pre-factor $S(\delta)$. Here, ${\hat f}(k)= b^2/(b^2+k^2)$;
substituting this in the expression of $\phi(1,\lambda)$ and
performing the integral yields $\phi(1,\lambda)=
(b+\lambda)/\lambda$. Hence, ${\tilde h}(\lambda)=
(b+\lambda)/\lambda^2$. This Laplace transform can be readily
inverted to give $h(z)=1+bz$. Using this explicit form of $h(z)$
in the expression for $S(\delta)$ in (\ref{rnavg.2}) and
performing the integral yields an exact expression for the
pre-factor, valid for all $\delta\ge 0$
\begin{equation}
S(\delta)= \frac{2}{\sqrt{\pi}}\, \exp\left[-b\, \delta\right]\, .
\label{exp.1}
\end{equation}
Note that when $\delta\to 0$, one recovers the universal pre-factor $2/\sqrt{\pi}$.

\par
Consider next a jump density, $f(\eta)$, whose tail decays as
$f(\eta)\sim \exp[-|\eta|^{a}]$ for large $\eta$, where $a>0$.
Substituting this in the expression for $S(\delta)$ in
(\ref{rnavg.2}), expanding for large $\delta$ and using $h(0)=1$,
one can show that for large $\delta$,
$S(\delta) \sim \delta^{1-a}\, e^{-\delta^{a}}$.
For example, for the Gaussian distribution, $f(\eta)=
e^{-\eta^2/{2\sigma^2}}/{\sqrt{2\pi \sigma^2}}$, one finds that
\begin{equation}
S(\delta)\xrightarrow[\delta\to \infty]{} \frac{\sqrt{2}}{\pi}\,
\frac{\sigma}{\delta}\, e^{-\delta^2/{2\sigma^2}}\, .
\label{gaussian.1}
\end{equation}

\par
Finally, consider jump densities with power law tails,
$f(\eta)\sim |\eta|^{-\mu-1}$ for large $\eta$ with $\mu>0$. For
L{\'e}vy flights, $0<\mu<2$, whereas for jump densities with a
finite variance, $\mu\ge 2$. The Fourier transform, ${\hat f}(k)$,
for small $k$, generically behaves as
\begin{equation}
{\hat f}(k) \xrightarrow[k\to 0]{} 1- |ak|^{\mu} + O(k^2).
\label{fk.1}
\end{equation}
In this case, for large $\delta$ the dominant contribution to the
integral $S(\delta)=(2/\sqrt{\pi})\, \int_0^{\infty} h(z)
f(z+\delta) dz$ comes from the large $z$ region. For large $z$,
one can show that $h(z)$ has the asymptotic behavior
\begin{eqnarray}
h(z) &\approx & \frac{1}{a^{\mu/2}\, \Gamma[1+\mu/2]}\, z^{\mu/2} \quad {\rm for}\,\, \mu<2 \label{mlt2} \\
&\approx & \frac{\sqrt{2}}{\sigma}\, z \quad {\rm for}\,\, \mu\ge
2 . \label{hz.1}
\end{eqnarray}
Now, consider $S(\delta)= (2/\sqrt{\pi})\, \int_0^{\infty}
f(z+\delta)\, h(z)\, dz$. We first rescale $z=\delta y$. This
gives $S(\delta)= (2/\sqrt{\pi})\, \delta\, \int_0^{\infty}
f(\delta(y+1))\, h(y\delta)\, dy$. For large $\delta$, we use
(\ref{hz.1}) to obtain,
\begin{equation}
S(\delta)\xrightarrow[\delta\to \infty]{} \sim
\delta^{-\mu+\alpha} \label{sd.1}
\end{equation}
where $\alpha= \mu/2$ for $\mu\le 2$ and $\alpha=1$ for $\mu\ge
2$. Thus, in this case $S(\delta)$ decays as a power law for large
$\delta$.

\par
To test these analytical predictions we have performed Monte Carlo
simulations for the three jump densities: (i) $f(\eta)= (1/2)\,
\exp[-|\eta|]$ (Exponential); (ii) $f(\eta)= (1/\sqrt{2\pi})\,
\exp[-\eta^2/2]$ (Gaussian), and (iii) $f(\eta)$ drawn from a
L\'evy distribution with L\'evy exponent $\mu=1$. The L{\'e}vy
random number was generated using the method of \cite{Weron} and
\cite{Chambers}. While (i) and (ii) represent normal Fickian
diffusion, the L\'evy case represents non-Fickian (anomalous)
diffusion; the latter can arise in diverse heterogeneous domains
such as cells \cite{GC06}, cold atoms \cite{Sagi12}, and
disordered porous media \cite{DCSB04,ESB10}.

\par
\begin{figure}
\includegraphics[scale = 0.50] {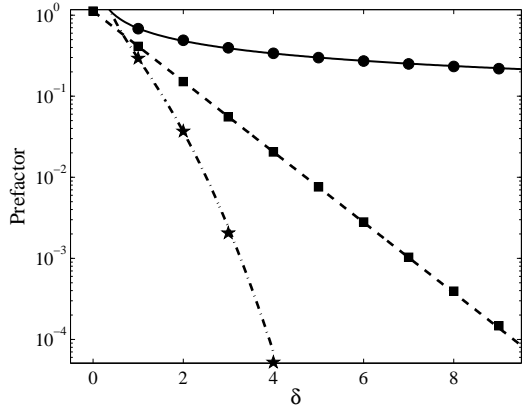}
\caption{One-dimensional pre-factor $S(\delta)$ versus measurement
error $\delta$ with Gaussian (stars), exponential (squares; $b=1$)
and L{\'e}vy (circles; $\mu=1$) jump length pdf's. The curves
(dotted-dashed, dashed, solid) are the corresponding analytical
results from (\ref{gaussian.1}), (\ref{exp.1}) and (\ref{sd.1})
with, respectively, functional forms $\frac{\sqrt{2}}{\pi\,
\delta}\,\exp[-\delta^2/2]$, $(2/\pi^{1/2})\,\exp(-\delta)$, and
$0.69\, \delta^{-0.51}$. In the L\'evy case, $\mu=1$, hence
$\alpha=\mu/2=1/2$, and the theoretical prediction $\sim
\delta^{-1/2}$ in Eq. (\ref{sd.1}) is consistent with
simulations.}
%
\label{Fig1}
\end{figure}

Our simulations are conducted with an ensemble of independent
random walkers,  each entering the one-dimensional system at the
origin, and with the jump length at each step drawn independently
from a given pdf. In every simulation, 5000 particles take $10^6$
steps each. In all cases, the particle is moved from step to step
according to its actual (sampled) location, without including the
value of $\delta$; $\delta$ is added as a fixed fraction of the
mean (median, for the L{\'e}vy pdf) jump length. At each step, the
location of the particle is calculated and the current distance
value must exceed the last maximum by at least the measurement
error $\delta$ to qualify as a new $\delta$-record and be counted;
otherwise we ignore it. The simulations confirm the $n^{1/2}$
scaling for the growth of mean number of $\delta$-records, for all
values of $\delta$. Furthermore, the three analytical predictions
for $S(\delta)$ in (\ref{exp.1}), (\ref{gaussian.1}) and
(\ref{sd.1}) are compared to Monte Carlo simulations in Fig.
\ref{Fig1}; the agreement is excellent. The pre-factor $S(\delta)$
decreases from its universal value $S(0)=2/{\sqrt{\pi}}$ as
$\delta$ increases, so that fewer records are counted as the error
increases. It is seen that the decrease in $S(\delta)$ is steepest
for the Gaussian pdf and has a much slower decay for the L{\'e}vy
pdf, in complete agreement with theory. The slowing down in the
L\'evy case is due to the anomalously skewed nature of the pdf,
with frequent small jumps and rare but enormous leaps; as a
consequence, potential records set by small jumps are more prone
to being eliminated by the $\delta$ error. In contrast,  the
Gaussian case with $\sigma=1$ displays a rapid decline with the
increasing error, due to the compactness of the pdf, so that large
jumps are quite rare and record events larger than the error are
rarer yet.

\par
\begin{figure}
\includegraphics[scale = 0.50]{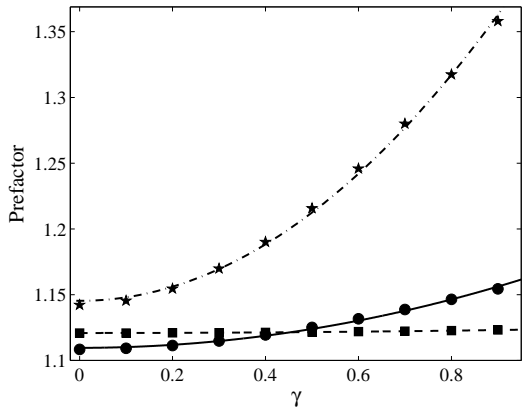}
\caption{Pre-factor $T(\gamma)$ as a function of the measurement
noise $\gamma$ for jump lengths (in one dimension) with Gaussian
(stars), exponential (squares; $b=1$) and L{\'e}vy (circles;
$\mu=1$) pdf's.
The curves represent quadratic fits of functional form $a+b
\gamma^2$ with different values of $a$, $b$.
%
} \label{Fig2}
\end{figure}

We now proceed to examine the influence of the measurement noise
$\gamma$. Let $\{x_0=0,x_1,x_2,\ldots, x_n\}$ represent the
successive positions of the random walker. In this case, a record
is registered at step $m$ if
\begin{equation}
x_m+ {\cal N}(0,\gamma)\, \Delta x > {\rm max}(0,x_0,x_1,\ldots,x_{m-1})
\label{gamma.1}
\end{equation}
where ${\cal N}(0,\gamma)$ is a zero-mean Gaussian random variable
with a standard deviation $\gamma$. The characteristic magnitude
of the jump length, $\Delta x$, is chosen as $\Delta x=1$ for the
exponential pdf (i) and $\Delta x= \sigma=1$ for the Gaussian pdf
(ii); for the L\'evy pdf (iii), $\Delta x$ is the median of the
one-sided L\'evy distribution with $\mu=1$. The term ${\cal
N}(0,\gamma)\, \Delta x$ in (\ref{gamma.1}) mimics the measurement
noise. The noise is added for the purpose of record verification
at each step and is not accumulated to the actual sequence. An
analytical treatment analogous to that for $\delta$ is not yet
available and we resort to numerical experiments, similar to those
for $\delta$, with the results shown in Fig. \ref{Fig2}.

\par
While the scaling $\langle R_n\rangle \sim T(\gamma)\, n^{1/2}$
for large $n$ persists, in stark contrast to the $S(\delta)$, the
pre-factor $T(\gamma)$ shown in Fig. \ref{Fig2} is an increasing
function of $\gamma$ for all jump densities.
Thus for $\gamma$-records, the noise adds spuriously to the
record-setting events, leading to false accounting of records and
rendering an \textit{apparent} $\langle R_n\rangle$ larger than
the actual one. This spuriously large rate of record formation
increases with the magnitude of the noise and suggests that it
might be possible to infer ``signal-to-noise'' ratio in
diffusion-type experiments by means of record counting.

\par
For $\gamma$-records, the universality of record-setting affords
the opportunity to estimate, a priori, the magnitude of $\gamma$
in an experiment involving an ensemble of measurements. At least
in principle, one first determines from an experiment the pdf of
the jump lengths in the domain. This pdf can then be employed in
random walk simulations, as shown above, to generate a curve for
the pre-factor $T(\gamma)$ (such as seen in Fig. \ref{Fig2}).
Returning then to an ensemble of experimental measurements in the
real system, one determines $T$ and then reads off the
corresponding value of $\gamma$ from the simulated $T(\gamma)$
curve. This may provide a practical and simple algorithm to
estimate the measurement noise $\gamma$ in a given experimental
setup.

\par
The results presented here illustrate the subtlety and richness of
record-breaking and counting, in the presence of instrumental
error $\delta$ and measurement noise $\gamma$, in systems where
the underlying process can be modelled by a random walk. The
decoupling of the growth exponent ($1/2$, regardless of precision
and noise) from the pre-factor (which depends on instrumental
precision and noise in a monotonic, contrasting, and pdf-dependent
manner) is significant. While the universality of the mean record
number persists, $\langle R_n\rangle \sim n^{1/2}$, the magnitude
of the pre-factor is sensitive to the presence of $\delta$ or
$\gamma$. Such sensitivity can, perhaps, be exploited in real
experiments to infer instrumental uncertainty and noise from
record-counting data.

\par
Finally, we note that all the above Monte Carlo simulations were
also performed on 2d and 3d orthogonal lattices. The universality
of the $n^{1/2}$ record-setting scaling is robust for all
dimensions, and in all cases, the pre-factors displayed
qualitative behaviors similar to those shown in Figs. \ref{Fig1}
and \ref{Fig2}. Moreover, Monte Carlo simulations accounting for
two-sided records (absolute distance) demonstrated the same
$n^{1/2}$ and similar qualitative behavior for the dependence of
the pre-factors on $\delta$ and $\gamma$.

B.B. acknowledges support from the Israel Science Foundation
(Grant No. 221/11). A.B.K. acknowledges NSF grant AGS-1119164 and
hospitality of the Weizmann Institute of Science. S.N.M.
acknowledges ANR grant 2011-BS04-013-01 WALKMAT. Part of this work
was carried out while S.N.M. was a Weston Visiting Professor at
the Weizmann Institute of Science.


\begin{thebibliography}{99}

\bibitem{FS54} F. G. Foster and A. Stuart, J. Roy. Stat. Soc. {\bf 16}, 1 (1954).

\bibitem{Arnold} B. C. Arnold, N. Balakrishnan, and H. N. Nagaraja, {\it Records}, Wiley (1998).

\bibitem{Nevzorov} V. B. Nevzorov, {\it Records: Mathematical Theory}, Am.
Math. Soc. (2001).

\bibitem{Glick78} N. Glick, Am. Math. Mon. \textbf{85}, 2 (1978).

\bibitem{Gembris2002} D. Gembris, J. G. Taylor, and D. Suter, Nature \textbf{417}, 506 (2002).

\bibitem{Gembris2007} D. Gembris, J. G. Taylor, and D. Suter, J. Appl. Stat. \textbf{34}, 529 (2007).

\bibitem{BRV07} E. Ben-Naim, S. Redner, and F. Vazquez, Europhys. Lett.
77, 30005 (2007).

\bibitem{KJ05} J. Krug and K. Jain, Physica A {\bf 358}, 1 (2005).

\bibitem{Krug07} J. Krug, J. Stat. Mech. P07001 (2007).

\bibitem{Oliveira2005} L. P. Oliveira, H. J. Jensen, M. Nicodemi, and P. Sibani, Phys. Rev. B \textbf{71}, 104526 (2005).

\bibitem{Sibani2006} P. Sibani, G. F. Rodriguez, and G. G. Kenning, Phys. Rev. B \textbf{74}, 224407 (2006).

\bibitem{GL08} C. Godr{\`e}che and J. M. Luck, J. Stat. Mech. P11006 (2008).

\bibitem{Matalas97}
N. C. Matalas, Climatic Change {\bf 37}, 89 (1997); R. M.
Vogel, A. Zafirakou-Koulouris, and N. C. Matalas, Water
Resour. Res. {\bf 37}, 1723 (2001).

\bibitem{Redner2006} S. Redner and M.R. Petersen,  Phys. Rev. E \textbf{74}, 061114 (2006).

\bibitem{Meehl2009} G. A. Meehl, C. Tebaldi, G. Walton, D. Easterling, and L. McDaniel,
Geophys. Res. Lett. \textbf{36}, L23701 (2009).

\bibitem{WK10} G. Wergen and J. Krug, Europhys. Lett. {\bf 92}, 30008 (2010).

\bibitem{AK11} A. Anderson and A. Kostinski, J. Appl. Meteo. and Climat. {\bf 50}, 1859
(2011).

\bibitem{SLJ13} S. C. L. Srivastava, A. Lakshminarayan, and S. R. Jain, Europhys. Lett. {\bf 101}, 10003 (2013).

\bibitem{Williams06} R. J. Williams, {\it Introduction to the Mathematics of Finance} (AMS, Providence, 2006).

\bibitem{Yor2000} M. Yor, {\it Exponential Functionals of Brownian motion and Related Topics} (Berlin, Springer, 2000).

\bibitem{Asmussen} S. Asmussen, Applied Probability and Queues (Springer,
New York, 2003); M. J. Kearney, J. Phys. A {\bf 37}, 8421
(2004).

\bibitem{MZ08} S. N. Majumdar and R. M. Ziff, Phys. Rev. Lett.
{\bf 101}, 050601 (2008).

\bibitem{EKB11} Y. Edery, A. Kostinski, and B. Berkowitz, Geophys. Res. Lett. {\bf 38},
L16403 (2011).

\bibitem{PLDW09} P. Le Doussal and K. J. Wiese, Phys. Rev. E {\bf 79}, 051105 (2009).

\bibitem{MSW12} S. N. Majumdar, G. Schehr, and G. Wergen, J. Phys. A: Math. Theor. \textbf{45}, 355002 (2012).

\bibitem{SS} S. Sabhapandit, Europhys. Lett. {\bf 94}, 20003 (2011).

\bibitem{WMS12} G. Wergen, S. N. Majumdar, and G. Schehr, Phys. Rev. E {\bf 86}, 011119 (2012).

\bibitem{AK10} A. Anderson and A. B. Kostinski, J. Appl. Meteor. Climatol.
49, 1681 (2010).

\bibitem{WVRK12} G. Wergen, D. Volovik, S. Redner, and J. Krug, Phys. Rev. Lett. {\bf 109}, 164102 (2012).

\bibitem{BBP96} N. Balakrishnan, K. Balasubramanian, and S. Panchapakesan,
J. Appl. Stat. Sci. 4, 123 (1996).

\bibitem{GJS12} R. Gouet, F. J. Lopez, and G. Sanz, Comm. Stat. Theor. Methods {\bf 41}, 309 (2012).

\bibitem{SMreview} S. N. Majumdar, Physica A 389, 4299 (2010).

\bibitem{Pollaczek} F. Pollaczek, Comptes rendus 234, 2334 (1952).

\bibitem{Spitzer} F. Spitzer, Duke Math. J. 24, 327 (1957).

\bibitem{SA} E. Sparre Andersen, Matematica Scandinavica 2, 195 (1954).

\bibitem{Weron} A. Weron and R. Weron, Lec. Notes in Physics 457, 379 (1995).

\bibitem{Chambers} J. M. Chambers, C. L. Mallows, and B. W. Stuck, JASA 71, 340 (1976).


\bibitem{GC06} I. Golding and E. C. Cox, Phys. Rev. Lett. 96, 098102 (2006).

\bibitem{Sagi12} Y. Sagi, M. Brook, I. Almog, and N. Davidson, Phys. Rev. Lett.
108, 093002 (2012).

\bibitem{DCSB04} M. Dentz, A. Cortis, H. Scher, and B. Berkowitz, Adv. Water
Resour. 27, 155 (2004).

\bibitem{ESB10} Y. Edery, H. Scher, and B. Berkowitz, Water Resour. Res. 46,
W07524 (2010).

\end{thebibliography}

\end{document}